\documentclass[acmlarge]{acmart}
\makeatletter
\newcommand{\confshort}{\acmConference@shortname}
\newcommand{\conffull}{\acmConference@name}
\newcommand{\confdate}{\acmConference@date}
\newcommand{\confloc}{\acmConference@venue}
\AtBeginDocument{
  \fancypagestyle{firstpagestyle}{
    \fancyhead{}%
    \fancyfoot[C]{}%
  }
  \fancyhf{}
  \fancyhead[LO]{\@headfootfont\shorttitle}%
  \fancyhead[RE]{\@headfootfont\@shortauthors}%
  \fancyhead[LE]{\@headfootfont\footnotesize \confshort, \confdate, \confloc}%
  \fancyhead[RO]{\@headfootfont\footnotesize \confshort, \confdate, \confloc}%
  \fancyfoot[C]{}%
}
\makeatother
\acmBooktitle{\conffull\@ (\confshort), \confdate, \confloc}
\AtBeginDocument{%
  }

\usepackage{longtable}
\usepackage{booktabs}
\usepackage{array}
    \newcolumntype{R}{>{\raggedleft\arraybackslash}p{2cm}}
\usepackage{tcolorbox}
\copyrightyear{2026}
\acmYear{2026}
\setcopyright{cc}
\setcctype{by}
\acmConference[FAccT '26]{The 2026 ACM Conference on Fairness, Accountability, and Transparency}{June 25--28, 2026}{Montreal, QC, Canada}
\acmBooktitle{The 2026 ACM Conference on Fairness, Accountability, and Transparency (FAccT '26), June 25--28, 2026, Montreal, QC, Canada}
\acmDOI{10.1145/3805689.3812399}
\acmISBN{979-8-4007-2596-8/2026/06}

\begin{document}

\title[Strategic Polysemy in AI Discourse]{Strategic Polysemy in AI Discourse: A Philosophical Analysis of Language, Hype, and Power}
\author{Travis LaCroix}
\affiliation{%
  \institution{Durham University}
  \city{Durham}
  \country{UK}
}
\email{travis.lacroix@durham.ac.uk}

\author{Fintan Mallory}
\affiliation{%
  \institution{Durham University}
  \city{Durham}
  \country{UK}
}
\email{fintan.d.mallory@durham.ac.uk}

\author{Sasha Luccioni}
\affiliation{%
  \institution{Hugging Face}
  \city{Montr{\'e}al}
  \country{Qu{\'e}bec}
}
\email{sasha.luccioni@huggingface.co}

\renewcommand{\shortauthors}{LaCroix et al.}

\begin{abstract}
        This paper examines the strategic use of language in contemporary artificial intelligence (AI) discourse, focusing on the widespread adoption of metaphorical or colloquial terms like ``hallucination'', ``chain-of-thought'', ``introspection'', ``language model'', ``alignment'', and ``agent''. We argue that many such terms exhibit {\it strategic polysemy}: they sustain multiple interpretations simultaneously, combining narrow technical definitions with broader anthropomorphic or common-sense associations. In contemporary AI research and deployment contexts, this semantic flexibility produces significant institutional and discursive effects, shaping how AI systems are understood by researchers, policymakers, funders, and the public. To analyse this phenomenon, we introduce the concept of {\it glosslighting}: the practice of using technically redefined terms to evoke intuitive---often anthropomorphic or misleading---associations while preserving plausible deniability through restricted technical definitions. Glosslighting enables actors to benefit from the persuasive force of familiar language while maintaining the ability to retreat to narrower definitions when challenged. We argue that this practice contributes to AI hype cycles, facilitates the mobilisation of investment and institutional support, and influences public and policy perceptions of AI systems, while often deflecting epistemic and ethical scrutiny. By examining the linguistic dynamics of glosslighting and strategic polysemy, the paper highlights how language itself functions as a sociotechnical mechanism shaping the development and governance of AI. 
\end{abstract}



\received{13 January 2026}
\received[revised]{25 March 2026}
\received[accepted]{16 April 2026}


\maketitle

\section{Introduction}
\label{sec:Introduction}

    ``Artificial intelligence'' (AI) was coined by John McCarthy in 1955 as a provocative label for a research programme intended to attract attention and secure funding for the Dartmouth Summer Workshop \citep{McCarthy-et-al-1955}. The choice of terminology was {\it strategic}: it distinguished McCarthy's approach to symbolic logic and computational reasoning from Norbert Wiener's {\it Cybernetics} programme \citep{Wiener-1948}, which studied agents, feedback, and control in their interaction with the environment. Yet the idea of AI long predates this coinage. If ``artificial intelligence'' is taken to mean non-biological or non-natural intelligent systems---artifacts of engineering, religion, or magic---then the concept dates back (at least) to the ancient Greeks. The idea of creating artificial life from inanimate matter is present in creation stories from Sumerian, Chinese, Jewish, Christian, and Muslim traditions \citep{Mayor-2018, Coeckelbergh-2020}. Moreover, both symbolic (logic-based) and sub-symbolic (machine learning) approaches to computational systems existed well before the term ``AI'' entered the lexicon.%
        \footnote{For example, \citet{McCulloch-Pitts-1943} proposed a mathematical model of an artificial neuron, drawing on contemporary knowledge of the basic physiology and functioning of biological neurons, propositional logic, and Turing's theory of computation. See discussion in \citet[Ch. 1]{LaCroix-2025}.} 

    These historical and fictional precursors highlight a conceptual and linguistic problem. ``Artificial intelligence'', in the sense that it is advertised today---i.e., as a current or impending reality---does not exist. Scholars have highlighted that what we {\it call} ``artificial intelligence'' is neither artificial nor intelligent \citep{Broussard-2018, Crawford-2021}. This is not to say that ``artificial intelligence'' is a compositional contradiction---like, say, ``progressive conservative''. Nor is it that the denoting phrase expressed by ``artificial intelligence'' fails to denote---like ``the present King of France''. Instead, there has always been a gap between the colloquial meaning of ``intelligence'' and the technical reality of the systems the term is used to describe. Crucially, this gap is not a recent divergence introduced by deep learning or large language models (LLMs); it is a structural feature of AI research that has existed since its very inception.

    From the outset of the field, then, the language of artificial intelligence has been inseparable from persuasion, speculation, and strategic self-presentation. The term ``AI'' has always functioned as much as a rhetorical device as a technical descriptor, mobilising metaphors of human cognition and creativity to attract attention, resources, and legitimacy. Over subsequent decades, this tendency has only intensified: new subfields, funding cycles, and public narratives have continually repackaged computational methods in linguistically inflated or anthropomorphic terms. The history of AI is not only a history of technological development but also a history of terminological opportunism, in which language plays an active role in producing and sustaining the very phenomena it purports to describe \citep{edwards1996closed}.

    To understand how this linguistic opportunism operates today, we need to examine the mechanisms through which language in AI terminology does its conceptual and rhetorical work. As such, this paper has two key aims. First, it distinguishes between two kinds of linguistic ambiguity that structure contemporary AI discourse. {\it Innocent polysemy} refers to the ordinary, context-sensitive flexibility of meaning characteristic of natural language. In contrast, {\it strategic polysemy} involves the (either deliberate or merely foreseeable) exploitation of that flexibility for rhetorical, financial, or ideological gain. We suggest that strategic polysemy pervades AI research communication---particularly through anthropomorphic, sensational, or hype-laden terminology. Second, we argue that strategic polysemy enables a distinctive mode of epistemic manipulation we call {\it glosslighting}: the use of common and colloquially well-understood terms applied in a narrow, technical sense to evoke familiar, anthropomorphic meanings while preserving plausible deniability. Through glosslighting, researchers and corporations benefit from public enthusiasm while shielding themselves from epistemic, moral, and (possibly) legal accountability.

    The guiding issue is therefore both conceptual and normative: it concerns the consequences that arise when a technical field systematically adopts misleading language to describe its products and processes. In what follows, we argue that these consequences are neither incidental nor benign---strategic ambiguity in AI terminology drives hype cycles, obscures limitations, and shapes policy and public understanding in normatively meaningful ways.

    In what follows, Section~\ref{sec:Polysemy} develops our conceptual foundation, drawing on the philosophy of language to distinguish ``innocent'' and ``strategic'' polysemy. Section~\ref{sec:Glosslighting} applies this framework to contemporary AI discourse, examining a series of increasingly common and influential terms---e.g., {\it agent}, {\it hallucination}, {\it intelligence}, {\it reasoning}, {\it alignment}, etc.---to demonstrate how the superposition of colloquial and technical senses produces what we call {\it glosslighting}. Section~\ref{sec:Hype} situates glosslighting within the broader political economy of AI, arguing that anthropomorphism and strategic polysemy are central drivers of the modern AI hype cycle and tracing how corporate, academic, media, and investment ecosystems amplify these linguistic choices. Section~\ref{sec:Normativity} turns to the normative implications of glosslighting behaviour, showing how it generates epistemic, ethical, and structural harms, and proposing a framework for responsible terminological practice. 

\section{Ambiguity and Polysemy in Natural Language}
    \label{sec:Polysemy}

    Polysemy is a linguistic phenomenon wherein a single word form is associated with two or more related meanings (or senses) \citep{Falkum-Vincente-2015}. The {\it relatedness} of the senses of a polysemous word form will be particularly important: this feature distinguishes polysemy from, e.g., homonymy, wherein two {\it unrelated} meanings happen to share a word form, as in ``ball'', which can refer to both a physical object and event \citep{Liu-2025}.

    Innocent polysemy is an inevitable feature of natural language, and a speaker need not necessarily intend that a word be interpreted in accordance with a single, maximally precise sense in order to communicate. In contrast, strategic polysemy occurs when a speaker does intend to communicate distinct senses of a word, either to different audiences, or to the same audience while preserving the deniability of at least one of those senses. As we describe below, strategic polysemy need not be malicious---for example, puns depend upon strategic polysemy for humorous purpose.
    
    The property of deniability gives strategic polysemy some overlap with {\it insinuation} \citep{Camp-2018}. In cases of insinuation, a speaker strategically crafts an utterance in such a way so as to minimise conversational risk, ensuring that their explicit on-record content is unobjectionable while their riskier conversational move is implicit. When called out by a pedant, the insinuating speaker can deny they meant the riskier content. Camp models insinuation by distinguishing the common ground \citep{Stalnaker-2002, Stalnaker-2014}---determined by the mutual beliefs of speakers---from the conversational record \citep{Lewis-1979}---determined by what they are ``officially'' committed to. For example, when George W. Bush%
        \footnote{The 43rd president of the USA.}
    would reference the Dredd Scott case as an instance of an incorrect decision by the Supreme Court, he insinuated that he would be willing to overturn {\it Roe v. Wade} while not making any claims on the conversational record. Similar processes occur in cases of ``sneaky reference'' when a speaker knowingly uses referential terms in a way that they expect or intend to be misunderstood \citep{Michaelson-2022}. 

    What this body of literature from the philosophy of language should make clear is that polysemy, insinuation, dogwhistles, etc. arise from ordinary features of natural language and social interaction. However, these mechanisms take on distinctive forms---and acquire distinctive normative significance---when they are embedded in technical discourse that also functions as public communication. In what follows, we argue that contemporary discourse about artificial intelligence provides a salient case. Here, ambiguity and polysemy are not merely incidental by-products of language use, but are often systematically cultivated at the interface between technical research, corporate communication, media representation, and policy debate. Examining this domain allows us to see how familiar linguistic tools can be repurposed to shape epistemic expectations, manage accountability, and influence or structure public understanding, setting the stage for a specific type of strategic polysemy that we call ``glosslighting''. 

\section{Glosslighting: Ambiguity and Polysemy in AI Discourse}
    \label{sec:Glosslighting}

    As noted, not all ambiguity in language is deceptive. Strategic (also called purposeful or intentional) polysemy can be useful in helping the intended audience better understand the message that the speaker intends to transmit. For instance, when the speaker and receiver share a common background or frame of reference, they can use polysemy to strengthen social bonds or to explore new topics via a familiar lens that they both share~\citep{nerlich2001ambiguities}. Strategic polysemy is also commonly used in marketing to target a specific audience without explicitly referring to them---for instance, appealing to minority groups (e.g., LGBTQ+ or Black audiences) without losing consumers from the majority with the messaging~\cite{stefano2011two,tsai2012political}. The use of deliberate polysemy or ambiguity can also allow companies to have plausible deniability, making it a useful tool from a legal perspective~\cite{selivanovskikh2025strategic,hemel2023polysemy}. Importantly, our use of the term ``strategic'' does not require that speakers consciously intend to deceive. Rather, we use the term to refer to the foreseeable discursive effects of ambiguous language, the institutional incentives that reward its uptake, and the structural communicative advantages that such ambiguity affords when presented to heterogeneous audiences. In this sense, polysemy can function strategically even when individual speakers sincerely endorse one interpretation of the terms they use. In this context, forms of strategic polysemy can be differentiated, not in terms of ambiguity, but rather function and asymmetry. Strategic polysemy can become pernicious when ambiguity is systematically exploited to create misleading impressions for some audiences, while allowing speakers to disavow those impressions by retreating to alternative descriptions---i.e., the plausible deniability that seems inherent to strategic polysemy.%
        \footnote{This is distinct from the notion of an {\it empty} or {\it floating} signifiers in semiotics and political theory \citep{Levi-Strauss-1987, Laclau-2005}. An empty signifier is one whose meaning becomes indeterminate or widely contestable, often functioning to unify disparate political demands. By contrast, strategic polysemy involves the concurrent availability of multiple conflicting but plausible interpretations of a term, which can be differentially activated across audiences or contexts. The mechanism we analyse therefore concerns the strategic exploitation of semantic flexibility rather than the semantic indeterminacy characteristic of empty signifiers. See discussion in \citet{suchman2023uncontroversial}.}
    Here, the relevant explanatory factor is not necessarily the psychological intentions of particular speakers, but the foreseeable consequences of deploying polysemous terminology in specific institutional and communicative contexts.

    As it happens, contemporary discourse around artificial intelligence provides a particularly fertile environment for potentially {\it manipulative} strategic polysemy because this discourse is saturated with neologisms that migrate rapidly from technical coinage to popular vocabulary. More often than not, these terms carry anthropomorphic overtones. Terms such as {\it agentic AI}, {\it agent}, {\it hallucination}, {\it foundation model}, {\it chain-of-thought}, and {\it generative AI} are now ubiquitous across research papers, marketing materials, media coverage, and policy discussions. A striking feature of this vocabulary is its {\it systematic polysemy}. Many core AI terms simultaneously invoke a rich, intuitive meaning---often anthropomorphic and drawn from psychology, cognitive science, neuroscience, philosophy, or everyday social understanding---while also possessing a technical description that is far narrower, operational, and explicitly non-anthropomorphic. This dual structure enables researchers and corporate actors to reap the rhetorical benefits of anthropomorphic interpretation while retaining the ability to deny those interpretations when challenged. By anthropomorphism, we mean the projection---whether intentional or inadvertent---of characteristically human attributes like intentions, motivations, emotions, or deliberative capacities onto non-human entities \citep{Epley-et-al-2007}. While these projections have well-documented cognitive and social roots \citep{Urquiza-Haas-Kotrschal-2015}, in the context of AI anthropomorphism they also serve a {\it promotional} function. In this sense, anthropomorphic descriptions of AI systems' capabilities operate as a mode of hype, insofar as they systematically misrepresent, distort, or exaggerate the capacities of computational systems \citep{Placani-2024}. The following examples illustrate how this pattern functions in practice.

\paragraph{Hallucination}

    The term ``hallucination'' traditionally denotes vivid perceptual experiences with psychological or clinical significance. Its adoption in the field of AI to describe erroneous or fabricated model outputs imports mentalistic connotations that are not warranted by the underlying mechanisms because these models do not possess perceptual systems or experiential states; they generate outputs through statistical token prediction conditioned on training data. From this perspective, describing erroneous outputs as hallucinations risks conflating fundamentally different phenomena and may obscure the underlying mechanism responsible for these errors \citep{Emsley-2023}. A language model that outputs a fictitious citation is not ``imagining'' or misperceiving; it is producing a statistically-likely sequence given its training distribution. Nonetheless, the anthropomorphic label has been widely taken up in both technical and popular discourse, supporting headlines like ``Anthropic CEO claims AI models hallucinate less than humans'' \citep{TechCrunch-2025-Hallucination}, or ``A.I. Is Getting More Powerful, but Its Hallucinations Are Getting Worse'' \citep{NYT-2025-Hallucination}. However, this framing remains conveniently reversible. When anthropomorphic interpretations become inconvenient, researchers retain the ability to dismiss the intuitive meaning by appealing to the term's technical reinterpretation as a form of stochastic error \citep{Maleki-et-al-2024}.

\paragraph{Thinking, Reasoning, and Chain-of-Thought}

    Cognitive terms such as {\it thinking} and {\it reasoning} have recently re-entered technical AI discourse through the popularisation of ``chain-of-thought'' methods. In \citet{Wei-et-al-2023}, ``reasoning'' corresponds to producing intermediate text prior to outputting the solution to a problem while a ``chain of thought'' is ``a series of intermediate natural language reasoning steps that lead to the final output''. They also acknowledge that ``although chain of thought emulates the thought processes of human reasoners, this does not answer whether the neural network is actually `reasoning,' which we leave as an open question''. Despite this caveat, the title of the paper drops the scare-quotes and affirms ``Chain-of-Thought Prompting Elicits Reasoning in Large Language Models''. In practice, chain-of-thought ``reasoning'' (sometimes called chain-of-thought ``prompting'') simply instructs the model to generate intermediate text prior to producing a final solution---i.e., the text that would precede an answer to a question. Although \citet{Wei-et-al-2023} acknowledge ambiguity around whether ``reasoning'' aptly describes these systems' internal processes, the terminology has been widely adopted in high-profile technical reports, including those for GPT-4 \citep{OpenAI-Technical-Report-2024} and DeepSeek R1 \citep{DeepSeekAI-2025}, where chain-of-thought is presented as a key contributor to improved ``reasoning capabilities''---i.e., alleged improvement to performance on standard benchmarks. More recent systems expose this framing directly through parameters such as \texttt{enable\_Thinking = True}, which, when toggled,
    induce longer and more elaborate (``deliberative'') responses \citep{Yang-et-al-2025}; see, e.g., \href{https://huggingface.co/Qwen/Qwen3-0.6B}{Qwen 3}. The cumulative effect is to frame verbose, intermediate text as evidence of cognitive processing, while retaining the ability to retreat to a purely operational description when challenged.

\paragraph{Introspection}

    The term {\it introspection} has undergone a similar transformation. In contemporary technical usage, it refers to instances where a model can predict or report aspects of its own ``behaviour'' (outputs) more accurately than an external model can. One influential formulation characterises introspection as ``acquiring knowledge that is not contained in or derived from training data but instead originates from internal states'' \citep{Binder-et-al-2024}. Formally, this description is operationalised as a scenario in which a model, $M_{1}$, correctly reports a fact, $F$, that a ``stronger'' language model, $M_{2}$, cannot recover, even when $M_{2}$ has access to $M_{1}$'s training data.

    It is worth noting that discussions of findings surrounding purported ``introspection'' routinely deploy scare quotes to gesture at mental-state vocabulary while simultaneously disavowing its literal implications. For example, \citet{Ji-An-et-al-2025} write that ``It has been shown that LLMs can `introspect' -- acquiring knowledge of internal states that originates solely from those states and not from training data'' (9). In some cases, mentalistic notions are defined in terms of other mentalistic notions. For example, \citet{Lindsey-2025} claims that ``a model demonstrates introspective awareness if it can describe some aspect of its internal state''. These rhetorical manoeuvres exemplify strategic polysemy: the conceptual force of psychological terminology is preserved, while the ontological commitments that such terminology ordinarily entails are carefully hedged. In a telling footnote, \citet{Ji-An-et-al-2025} suggest that they ``use anthropomorphic terms (e.g., thought, metacognition, deception) to describe LLM behavior and internal activations, without implying human-like neural mechanisms, consciousness, or philosophical equivalence between humans and LLMs'' (1).

\paragraph{Language Model}

    The use of the term ``language model'' also participates in this pattern. To non-specialists, the term suggests that these systems model language in a linguistically or semantically meaningful sense---capturing meaning, communicative intent, or social context. However, there remains a range of divergent scholarly opinion on whether LLMs capture anything like meaning as understood within traditional metasemantic theories, and it is an open question whether this is possible.%
        \footnote{See \citet{Bender-Hanna-2025,baggio2026referential,mandelkern2024language,borg2025llms,Mallory-2023,Mallory-2025,ostertag-2025-language,milliere2024language,LaCroix-2024} for a range of viewpoints.} %
    At the very least, the claim that LLMs produce meaningful text conflicts with widely-held Gricean assumptions in linguistics and the philosophy of language which emphasise the role of communicative intentions in linguistic meaning. \citet{Rogers-Luccioni-2024} also highlight that ``despite its ubiquity, this definition [of the term language model] is far from clear''; they go on to propose a more formal definition and recommend that the term be avoided in scientific papers in lieu of more technical terms that describe the underlying architecture.

\paragraph{Alignment}

    The phenomenon that we have been describing is not relegated to purported abilities of AI systems, but also appears in normative and policy contexts. For example, the term ``alignment'' exhibits a similar dual structure. In popular and policy discourse, it evokes moral harmony between AI systems and human values, intentions, or societal goals \citep{Christian-2020, Gabriel-2020, Russell-2019}. Technically, however, alignment is typically operationalised through narrow proxies---e.g., reward modelling, preference learning, or constraint satisfaction---that fall short of capturing the normative richness implied by the term \citep{LaCroix-2024, LaCroix-2025}. This gap allows developers to gesture toward ethical assurance while limiting their commitments to specific, often brittle, technical innovations or performance on inapt benchmarks \citep{LaCroix-2022, LaCroix-Luccioni-2025}. It is worth noting that what we now call ``value misalignment'' or a ``misaligned AI system'' describes a system that exhibits unintended behaviour. In the context of the symbolic systems approach to artificial intelligence, when a system did not do what the developer expects or intends, this would be framed as a defect in the system's logic or specification---i.e., a bug---rather than a philosophical or normative mismatch between human values and machine objectives \citep{LaCroix-2025}. This modern terminology reflects a shift in perspectives on and practical approaches to poorly specified AI systems---i.e., from debugging code before deploying a system widely to managing the behaviour of large-scale systems after deployment.

\paragraph{Agent and Agentic AI}

    In everyday and philosophical contexts, an {\it agent} is an autonomous, goal-directed entity, capable of deliberation, intention, and (sometimes) moral responsibility. In economics, the term ``agent'' often functions as basically synonymous with ``actor''. By contrast, technical AI usage of the term ``agent'' is typically described as a programme that maps inputs to outputs, possibly with a reward signal or iterative loop. This description does not entail autonomy, intention, or awareness. Yet industry descriptions routinely foreground the folk meaning of agency. For example, Google describes AI agents as ``software systems that use AI to {\it pursue goals} and complete tasks on behalf of users. They show {\it reasoning}, {\it planning}, and {\it memory} and {\it have a level of autonomy} to {\it make decisions}, {\it learn}, and {\it adapt}'' \citep{Google-Cloud}. Such descriptions readily support media narratives like ``Who's to Blame When AI Agents Screw Up?'' \citep{Wired-2025}, ``Google's AI agents will bring you the web now'' \citep{TechCrunch-2025-Agent}, ``AI Agents Are Actually Making Remote Work More Human'' \citep{Entrepreneur-2025}---all of which followed shortly after Google's highly publicised I/O 2025 ``agentic'' announcements \citep{Pichai-2025}.%
        \footnote{The point here is not that the term ``agent'' lacks a legitimate technical history in AI or computer science. The notion of an ``agent'' as a system that selects actions in response to environmental inputs is well established in fields such as reinforcement learning, multi-agent systems, robotics, and decision theory. Instead, the concern is semantic drift across registers. Our point is that the use of a well-established folk-psychological concept, like agency, {\it allows} for oscillation between minimal technical meanings and richer everyday connotations, which can (later) be used to (strategically) obscure the extent to which (present) systems remain designer-specified computational processes rather than, e.g., autonomous agents in the philosophical sense \citep{sep-agency}.}

    The term ``agentic AI'', often attributed to Andrew Ng \citep{NYT-Agentic}, exemplifies this dynamic. Its polysemy allows developers to invoke robust intuitions about autonomy and initiative while retaining plausible deniability: when pressed, ``agentic'' reduces to a minimal, technical description. In practice, an ``agentic'' system is simply a loop that, e.g., generates text, parses outputs, calls tools or APIs, stores intermediate states, and repeats until a stopping condition is met. (This plausible deniability is normatively significant when considering that, according to the non-AI definition, agents are typically considered morally responsible.) The appearance of initiative arises not from autonomy, but from repeated model calls embedded in well-established architectures---e.g., sequential decision systems, planning loops, reinforcement learning models, and finite-state controllers---now paired with requests to LLMs. What is new with so-called ``agentic AI'' is not {\it agency}, but the rebranding of these old architectures through the embedding of LLMs, whose stochastic outputs can be mistaken for spontaneity or intention (and hence, the appearance of agency), with minimal additional engineering~\cite{bender2021dangers}. The novelty, then, lies less in the underlying mechanisms than in relabelling them as {\it agentic}.

\phantom{a}

    Although we have focused on just a few instances of the same phenomenon here, many other terms have similar characteristics as those described above---e.g., {\it consciousness}, {\it foundation model}, {\it general intelligence}, {\it beliefs / preferences / goals}, {\it creativity}, {\it planning / deliberation}, {\it goal-directed behaviour}, {\it empathy / affective AI}, {\it trustworthiness / honesty}, {\it personality}, {\it scheming}, etc. (See Appendix~\ref{sec:Appendix}.) Across these examples, the same pattern recurs, highlighting how these terms serve a dual purpose. They signal one thing to the public, policymakers, and investors---typically the intuitive, anthropomorphic, or metaphorical meaning---while simultaneously retaining a narrow, technical interpretation to which experts can retreat when challenged. This systematic ambiguity enables, what we call, {\it glosslighting}.%
        \footnote{See also discussion in \citet{Rehak-2021}.}
        \begin{quote}
        {\bf Glosslight} (Definition):\\
            {\it Glosslighting}%
                \footnote{This word is a portmanteau of {\it glossary}, meaning a list of terms relating to a specific subject, and {\it gaslight}, meaning to manipulate someone into questioning their own sanity or powers of reasoning. It is important to note that there is general agreement among philosophers and psychologists that gaslighting need not be intentional or conscious \citep{abramson2014turning, kirk2023dilemmatic, podosky2021gaslighting} and it is highly unlikely that, where intentional, glosslighting is driven by the same psychological processes as gaslighting behaviour. Indeed, as we shall see below, very different incentives may be at play. It may also be understood as a combination of {\it gloss}, meaning to add luster, make shine; the task of advertising, and {\it light}, to illuminate, capturing the extent to which discourse is being framed by glossy advertising rather than colder more scientific considerations. Although glosslighting can be understood as the act of {\it glossing over} something, this interpretation does not settle whether epistemic manipulation is involved, whereas our definition underscores that the possibility of such manipulation is a core feature of glosslighting.} %
            (verb) is the practice of using technically redefined or polysemous terms to evoke familiar meanings---often emotionally or cognitively powerful ones---while preserving the ability to deny those meanings through retreat into specialised, context-bound reinterpretations.
        \end{quote}
    The rhetorical effects of glosslighting---suggesting familiar meanings while retaining deniability---may arise intentionally, but they can also emerge from (i.e., are a systemic outcome of) the foreseeable interaction between ambiguous terminology (allowing multiple interpretations), and heterogeneous audiences (different groups interpreting the same term differently), combined with institutional incentives (encouraging strategic ambiguity). 

    When the speaker deliberately targets a particular audience for uptake (who may not be the immediate audience), this can give rise to cases of dogwhistling \citep{Saul-2018, Saul-2024}. Like dogwhistles and unlike standard cases of insinuation, glosslighting can also be performed unintentionally. \citet{Saul-2024} distinguishes between both intentional and unintentional dogwhistles as follows: %
        \begin{quote}
            {\bf Intentional}: a term or speech act with (at least) two plausible interpretations, such that one of these violates some widespread norm, and is meant to be understood primarily by those who are comfortable with this norm violation; and one appears innocent, and is meant to be understood primarily by those who would not want to see the norm being violated.
            
            {\bf Unintentional}: a term or speech act with (at least) two plausible interpretations, one of which violates some widespread norm, and one of which doesn’t violate that norm, which is used by someone unaware of the norm-violating interpretation. (13).
        \end{quote} %
    For example, terms like ``welfare reform'', ``inner city'', and ``sex-based rights'' can come to be adopted by people unaware of their origins and original discursive functions. This is vitally important. A person may unintentionally use terminology that serves the original discursive function of the dogwhistle, and in doing so impose a particular frame onto an issue, without intending to do so. Just as one may distinguish the act of dogwhistling from a dogwhistle, the expression by which the act is carried out, one might distinguish the act of glosslighting from the glosslight, the term which illuminates further discussion.
    
    In the context of AI, glosslighting is a form of epistemic manipulation that exploits linguistic ambiguity under the guise of scientific rigour. As \citet{Watson-2019} observes, even basic cognitive verbs, such as ``think'', ``learn'', or ``infer'', oscillate between literal and metaphorical uses in AI discourse, often without clarification. This ambiguity is hardly accidental---anthropomorphism is analytically embedded in the very concept of {\it artificial intelligence} \citep[p. 692]{Placani-2024}. The notion of artificial {\it intelligence} itself invites associations with human-like qualities in a non-biological entity, meaning that this sort of anthropomorphism is inherent to the research field by title alone. The term ``intelligence'' gives rise to intuitions about common sense, effective learning, planning, reasoning, and handling complex information across natural and abstract domains. However, no consensus definition of intelligence exists.%
        \footnote{\citet{Legg-Hutter-2007} discuss more than $70$ distinct definitions. Importantly, the history of intelligence (as a concept) is bound up with eugenics and race ``science'', with the standard metrics of ability often being employed to serve an idealogical function rather than a scientific one \citep{Richardson-2017}.}

    Hence, the ambiguity surrounding core terms in AI is not simply a matter of imprecise definition within the discipline itself. Because such terms already carry strong everyday and anthropomorphic associations, their meaning is especially sensitive to shifts in audience and context. What may function as a narrow, technical placeholder in one setting can easily be interpreted in a much broader sense in another. This becomes particularly significant once AI discourse moves beyond specialist communities and into public, political, or commercial areas, where the background assumptions required to interpret these terms in their technical sense can no longer be taken for granted. As \citet{Rehak-2021} notes:
        \begin{quote}
            Given the complexity of current technology, only experts can understand such systems, yet only a small number of them actively and publicly take part in corrective political exchanges about technology. Especially in the field of artificial intelligence (AI) a wild jungle of problematic terms is in use. However, as long as discussions take place among AI specialists those terms function just as domain-specific technical vocabulary and no harm is done. But domain-specific language often diffuses into other fields and then easily loses its context, its specificity and its limitations. (88--89)
        \end{quote}
    Glosslighting extends this account since the issue is not {\it only} that meaning shifts as terms travel, but that this very shift can be structurally advantageous. The same expression can evoke expansive interpretations when such interpretations are rhetorically useful, while still permitting retreat into a narrow, technical definition when challenged. In this way, {\it strategic} ambiguity does not merely arise from loss of context; instead, it is a resource that allows claims to appear more substantial, more scientifically grounded, or more socially acceptable than a strictly technical reading would warrant. It is precisely this entanglement of anthropomorphism with strategic ambiguity (audience-dependent interpretation and institutional incentives) that makes the field so amenable to glosslighting behaviour---and, as we argue below, why the resulting linguistic practices warrant sustained conceptual, linguistic, and normative scrutiny.%
        \footnote{Note that, although our analysis focuses on glosslighting in the context of AI, this does not mean that glosslighting (or adjacent behaviour) is unique to the field. For example, the political right has co-opted, redefined, and popularised ``woke'' from a community-specific term describing being ``alert to racial or social discrimination and injustice'' to a pejorative, catch-all term for anything perceived as overly liberal, progressive, or an existential threat to traditional culture and values \citep{Rose-2020} while simultaneously erasing Black history, Black liberation movements, and social justice \citep{Kelley-1962, NAACP-2023}. This re-defined, populist term shares many characteristics of glosslighting. 
        }

\section{Anthropomorphisation and Hype in AI Discourse}
    \label{sec:Hype}

    The linguistic practices described in the previous section do not operate in isolation. Instead, they become embedded within broader social, political, and economic systems in which anthropomorphism, ambiguity, and hype mutually reinforce one another. Glosslighting is therefore not merely a matter of careless language or isolated rhetorical excess; it is sustained and amplified by institutional incentives that reward visibility, excitement, and perceived breakthroughs. In this section, we describe how anthropomorphism functions as a constitutive mechanism of AI hype, how hype becomes institutionalised, and why contemporary AI discourse is especially susceptible to this dynamic.

\subsection{Anthropomorphism as a Constitutive Element of Hype}

    Hype in the context of artificial intelligence is not merely a distortion layered on top of an otherwise neutral description of technology.%
        \footnote{It is worth noting that hype has a long historical precedent in AI research, and this build up of hype has also precipitated the various ``AI winters''. See, e.g., \citet{Lighthill-1973} and discussion in \citet[Ch. 1]{LaCroix-2025}. See also discussion of hype in the context of economic bubbles offered by \citet{Floridi-2024, Widder-Hicks-2024}.}
    As \citet[p. 691]{Placani-2024} argues, hype consists in ``the misrepresentation and over-inflation of AI capabilities and performance'', and crucially, to participate in hype means actively contributing to its creation. That is, the elements that make up hype in relation to a particular subject---e.g., media coverage, marketing copy, keynote talks, social media threads, technical blog posts, public discourse---do not merely reflect excitement; instead, they actively generate the attention, anticipation, and urgency that define hype as a social phenomenon. When a new AI tool is announced, exaggerated claims in news articles, viral social media posts, and enthusiastic endorsements from influencers do more than simply report on the technology---they actively generate hype. Without these elements, the heightened excitement and attention surrounding the product would be far less pronounced.
    
    Both \citet{Placani-2024} and \citet{Bender-Hanna-2025} emphasise that hype operates by blending fact and fantasy in ways that drive adoption and attention. In this context, glosslighting terms create a sense of indispensability, presenting technologies not merely as useful tools, but as inevitable solutions to pressing social, economic, or existential problems.%
        \footnote{In this context, glosslighting encodes a form of ``technochauvinism'' \citep{Broussard-2018}; see also the discussion of inevitability in \citet{bareis2022talking}.}
    Glosslighting is an especially effective form of strategic polysemy in this regard because it leverages deeply ingrained cognitive and cultural schemas for understanding intelligence, agency, and social interaction. However, the anthropomorphic language that constitutes glosslighting does not merely accompany hype; instead, it helps produce the very misperceptions on which hype depends. By framing AI systems as entities that can ``think'', ``reason'', ``understand'', or ``decide'', such language invites audiences to infer capacities that are genuinely awe-inducing,%
        \footnote{Similar to, e.g., a magic show.}
    but which exceed what the systems actually implement.

    Importantly, hype of this form is ethically non-neutral. As \citet{Watson-2019} argues, anthropomorphism in AI discourse shapes moral intuitions, responsibility attributions, and trust relationships. When systems are framed as {\it agents} rather than {\it tools}, failures are more easily naturalised, accountability is diffused, and structural decisions recede from view. Glosslighting language, therefore, plays a key role in deflecting critical scrutiny even as it amplifies confidence and excitement. \citet{Bender-Hanna-2025} argue that this phenomenon can produce significant cultural, economic, and social effects and tends to recur alongside ongoing innovation or research, perpetuating a cycle that shapes how technologies and ideas are received, adopted, and imagined. At the same time, however, hype serves different purposes for distinct stakeholders: ``entrepreneurs who have products to sell, researchers who have academic departments to fund, and zealots that would benefit from such a fiction being perpetuated'' \citep[p. 22]{Bender-Hanna-2025}. The convergence of these incentives ensure that anthropomorphic framings are repeatedly reproduced, even while their limitations are well understood by experts.%
        \footnote{Although, sometimes it is less clear whether researchers genuinely believe the anthropomorphism, as in the case of high-profile researchers coming to believe that their company's AI systems were becoming sentient \citep{de-Cosmo-2022}, leading figures in the field suggesting that next-token prediction may be sufficient for artificial general intelligence \citep{Bubeck-et-al-2023, Wang-et-al-2024}, or that these systems could pose an existential threat to humanity \citep{Anthropic-2025}.}

\subsection{Linguistic Choice and Institutional Force}

    Glosslighting via anthropomorphic language plays a central role in the process of hype generation because ``In virtue of the attribution of distinct human characteristics that misrepresent and exaggerate AI capabilities and performance, anthropomorphism in AI can be viewed as a constitutive part of hype'' \citep[p. 691]{Placani-2024}. By using terms that evoke human-like properties to describe computational systems, speakers compress complex technical realities into intuitively graspable narratives. These narratives travel easily across institutional boundaries: from conference papers to corporate blog posts, from press releases to headlines, from investor pitch decks to policy briefings. Once established, such terms circulate with remarkable speed and stability, becoming taken-for-granted descriptors of technological reality rather than contestable metaphors.%
        \footnote{As a salient example: consider how much background (assumptions, definitions, understanding, etc.) must be made explicit in order to meaningfully assert that ``it is false that LLMs hallucinate''. By contrast, the claim ``LLMs hallucinate'' is typically treated as unproblematic and requires little or no clarification---i.e., it has become {\it common knowledge}. In the latter case, when someone utters the statement ``LLMs hallucinate'', the conversational record \citep{Lewis-1979} is straightforwardly updated---i.e., this utterance signals a shared set of tacit assumptions, commitments, and questions under discussion. By comparison, there are at least two distinct contexts in which one might assert ``LLMs do not hallucinate''. In the default context---where ``LLMs hallucinate'' is taken for granted---the speaker will have said something false  (for example, due to a mistaken belief about the system's behaviour.) In an alternative context, the speaker will have said something true, but this is because they mean something different by ``hallucinate'' than the (now) standard context. This kind of re-description and popularisation is part of how glosslighting allows manipulation by introducing {\it new} polysemy to an existing term: the hype generated changes the conversational record.}
    This circulation is not accidental. Glosslighting phrases are rapidly adopted, in part, because they perform multiple functions simultaneously: for example, in academic venues, they signal novelty and conceptual contributions \citep{Sramek-Yatani-2025, Rehak-2025, bareis2022talking}; in corporate contexts, they differentiate products in crowded markets \citep{Narayanan-Kapoor-2024, Bender-Hanna-2025}; in media discourse, they support compelling stories (or fear-mongering) \citep{Ryazanov-et-al-2024, Nkoala-et-al-2026, Bender-Hanna-2025}; and, in investment settings, they suggest innovation and scalability \citep{Goncalves-2025, Widder-Hicks-2024, Floridi-2024}. Hence, these terms inherently shape what is fund-able, publishable, and newsworthy.%
        \footnote{Compare the following two news releases, both published in the {\it New York Times}, around $60$ years apart: ``The Navy has revealed the embryo of an electronic computer today that it expects will be able to walk, talk, see, write, reproduce itself and be conscious of its existence'' \citep{NYT-1958}. ``AlphaZero seemed to express insight. It played like no computer ever has, intuitively and beautifully, with a romantic, attacking style. It played gambits and took risks \ldots AlphaZero had the finesse of a virtuoso and the power of a machine. It was humankind's first glimpse of an awesome new kind of intelligence'' \citep{NYT-2018}.}
    In the latter context, media amplification loops intensify this effect \citep{ONeil-2016}. Once a term is introduced in industry or research settings, journalists may be warranted in uncritically adopting them---i.e., if these terms are in use in academic articles (or, at least, preprints), then they are justifiably taken as an industry standard. However, glosslighting terms appear purpose-built for popular media uptake insofar as their anthropomorphic framing fits neatly with the incentives of 24-hour news cycles, click-driven headlines, and the demand for easily narrativised stories.

    As a result, these terms travel smoothly from technical contexts into journalistic reporting, where their simplifications and implications are not only preserved but amplified by wholly predictable popular interpretation. Hence, when headlines compare AI ``hallucinations'' to human cognition, they do more than mislead; they {\it stabilise} particular framings of AI. These framings, in turn, feed back into corporate messaging and research agendas, closing a self-reinforcing loop. Venture capital further accelerates this dynamic because evocative language is not merely tolerated but actively rewarded in environments that prioritise rapid growth, first-mover advantages, and narrative coherence. Glosslighting terms make speculative futures feel concrete and urgent, exploiting fear-of-missing-out while obscuring unresolved technical limitations. As a result, hype becomes structurally embedded in the AI ecosystem rather than a removable excess. 

    Taken together, this analysis underscores that glosslighting and anthropomorphic hype are not descriptive idiosyncrasies of AI discourse, but socially efficacious practices that shape how technologies are perceived, valued, and acted upon. By stabilising particular framings of AI capabilities across institutional contexts, these linguistic choices do genuine rhetorical work---i.e., they influence funding priorities, regulatory agendas, public trust, and expectations about what AI systems are and what they ought to be able to do. This observation raises questions that cannot be answered at the level of descriptive analysis alone. Namely, if glosslighting language functions as an infrastructural component of the AI ecosystem, then its introduction and use cannot be normatively neutral. Hence, we turn now from diagnosis to evaluation, examining the epistemic, moral, and structural consequences of glosslighting in AI discourse.

\section{The Normative Dimensions of Language}
    \label{sec:Normativity}

    The use of strategic polysemy can have a multitude of consequences (intended or not) and result in many different kinds of harms when assumptions are made by the people on the receiving end of the discourse. Notably, in terms of epistemic harms, strategic polysemy can undermine the public understanding of an advanced technology whose workings are already impenetrable to the majority of its users, as well as policymakers and even practitioners. Strategic polysemy in this setting, therefore, poses distinctive epistemic risks: it can actively undermine public understanding of an advanced technology whose mechanism, limitations, and failure modes are difficult to scrutinise in the first place.

    When this epistemic gap is filled with glosslighting terms rather than careful, veridical explanation, the result is a distorted conception of AI systems' capabilities and limitations. Glosslighting terms actively {\it encourage} audiences to infer capacities that exceed the underlying technical reality, while obfuscating the contingent, brittle, and probabilistic nature of these systems. Over time, such distortions accumulate, shaping collective expectations about the abilities of these systems. One downstream effect of this distortion is the amplification of speculative narratives surrounding ``super-intelligence'' or artificial general intelligence (AGI). These thought experiments suppose that systems endowed with agency or intelligence that surpass humans could carry out actions that go against human well-being or survival---a common trope in science fiction---making popular audiences particularly susceptible to these narratives \citep{hermann2023artificial}. Importantly, these speculations do not arise in a vacuum; instead, they are scaffolded by the use of glosslighting terms. Hence, glosslighting contributes simultaneously to hype and to panic, inflating both utopian and dystopian interpretations of AI systems. In either case, the end-goal is the same: the generation of hype for the accumulation of capital. 

    Glosslighting also enables a subtler but equally harmful outcomes: it licenses pseudoscientific claims regarding AI systems' capabilities while avoiding accountability for the consequences of this discourse, since the claims that are made remain open to interpretation. Much like the use of strategic polysemy in marketing and political communication, the ambiguity allows speakers to gesture toward extraordinary claims without fully committing to them. As suggested above, when challenged, these claims can be reinterpreted, narrowed, or reframed as misunderstandings on the part of the audience rather than as overstatements by the speaker. This interpretive flexibility makes it hard to assign responsibility for the social, economic, or policy consequences of AI discourse on any one group within the AI community, even when that discourse demonstrably shapes public belief and institutional decision-making. 
    
    It is normal for a science to start with terms bearing potentially unclear or inexact everyday senses before precisifying them in ways that may not be completely satisfactory to the layperson \citep{carnap-1950}. Nevertheless, this process can be carried out more or less responsibly. For example, the public reception of Darwinian ideas was influenced by the use of terms such as ``fitness'', ``competition'', and ``superiority'' with devastating social consequences.%
        \footnote{See \citet{Ariew&Lewontin, lewontin1984not, gould1981mismeasure} for discussions of the history of social Darwinism.} %
    Knowingly using terms with normatively significant everyday senses to frame engineering developments carries with it considerable social risk which is not mitigated by the use of scare quotes, caveats, or footnotes clarifying that one doesn't mean the term in the everyday sense. While individual cases of explication or operationalisation may always be contested for failing to capture important aspects of the pre-theoretic conceptualisation or by loading them with unwarranted contents, operationalisations that risk seriously misleading the public and policy makers are best avoided.%
        \footnote{In his work on explication, \citet{carnap-1950} outlines several principles that should guide the process of transforming a pre-scientific concept into something scientifically rigorous; similarity to the original concept, exactness, fruitfulness and simplicity, as well as encouraging researchers to make explicit exactly which sense of the everyday concept they are seeking to capture; see also \citet{dutilh2020carnapian}. \citet{strawson1963carnap} objected to the Carnapian project of explication alleging that it merely involved ``changing the subject''. In the case of glosslighting, however, both the pre-scientific and the revisionary meanings of a term are exploited at the same time to make strong but deniable claims about novel technologies, a practice encouraged by the ``fake-it-'til-you make-it'' attitude among some AI researchers \citep{cummings2021rethinking}.}

    This illustrates that the harms associated with glosslighting terms (and their use) are not merely epistemic but also normative and structural. Epistemically, glosslighting degrades the quality of public knowledge about AI and prevents technological literacy surrounding increasingly ubiquitous systems. This is highly significant in light of recent studies that show receptivity to AI is predicted by lower AI literacy \citep{Tully-et-al-2025}. Hence, the epistemic harms associated with the use of glosslighting terms appear to be a feature from the perspective of marketing. Indeed, rather than advocating for AI literacy, one of the conclusions offered by \citet{Tully-et-al-2025} is that ``companies may benefit from shifting their marketing efforts and product development toward consumers with lower AI literacy''; moreover, ``efforts to demystify AI may inadvertently reduce its appeal'' \citep[p. 1]{Tully-et-al-2025}. (In this context, glosslighting might be seen as an effort to mystify AI.) Normatively, glosslighting manipulates trust by exploiting familiar concepts without respecting their connotations. Structurally, it reinforces power asymmetries between those who control the language used to describe AI---researchers, corporations, and institutions---and those who must rely on that language to make decisions about deployment, regulation, or adoption.

    Moreover, assigning responsibility for these dynamics is challenging, precisely because AI discourse is highly distributed and propagates rapidly across domains---from conference papers and technical reports to press releases, journalism, and social media, to common parlance. Nonetheless, as \citet{Rogers-Luccioni-2024} emphasise, researchers and developers training and deploying AI systems have a particularly important role to play in shaping the terms through which these systems are understood. Hence, they can opt to favour precise, literal, and technical descriptions over ambiguous or metaphor-laden ones, which are then left open to interpretation~\cite{Rogers-Luccioni-2024, Inie-et-al-2026}. This is especially pressing in contexts where their work is likely to be cited, popularised, or repurposed beyond expert audiences---i.e., those companies that have the highest concentration of power, market share, or influence.

    At the same time, scientific communicators and journalists also bear some responsibility for the interpretive frames they reproduce. Hence, they can choose to use terms that are more precise and less ambiguous. While accessible language is often necessary when communicating complex results, accessibility need not come at the expense of accuracy. Choosing terms that reflect the actual mechanisms underlying AI systems---even at the cost of rhetorical appeal---can help reduce the risk of misinterpretation, whether intentional or unintentional. Although some degree of sensationalism is constitutive of at least some part of journalistic discourse, especially in a novel and fast-paced field like AI, avoiding intentionally misleading claims regarding the capabilities and limitations of AI models can help damper both unwarranted hype and disproportionate panic surrounding AI systems.

\subsection{The Price of a Metaphor}

    It is worthwhile to note that not all metaphor or anthropomorphism in science is misguided or deceptive. Metaphorical language has long played a productive role in theory-building, pedagogy, and cross-disciplinary exchange \citep{Lakoff-Johnson-1980}. For example, the fluid dynamics of flowing water has been used to analogically reason about complex systems involving moving crowds or particles \citep{Gentner-Gentner-2014}. In the context of artificial intelligence, behaviourist work on reinforcement learning in the early 20th century influenced algorithms for temporal-difference learning; surprisingly, these algorithms then had a profound influence on neuroscience by providing a computational framework that explains the function of dopamine neurons as a reward prediction error signal \citep{Schultz-et-al-1997}. This feedback loop relied on shared metaphors---e.g., rewards, learning, policies---that facilitated mutual insight without collapsing distinctions between biological and artificial systems. In such cases, the metaphor can be understood as provisional, limited, and domain-specific.

    What distinguishes contemporary AI discourse is not that it uses metaphor, but rather, the context and uptake of the metaphors used. Up to the 1990s, anthropomorphic language circulated largely within expert communities that shared background assumptions about its limitations---i.e., the conversational ground and common record described above. Today, however, AI systems are widely deployed, widely discussed, and widely imagined as social actors by non-specialists. As a result, metaphors no longer remain inert explanatory devices; they actively shape public understanding, regulatory responses, and institutional trust. This shift creates a problem of model fit. Language that may be acceptable---or even useful---within a narrow research context can become misleading when it migrates into public discourse. In such settings, the ambiguity that once enabled productive abstraction now enables glosslighting. The rhetorical dynamics described here are not unique to AI. Techno-scientific history is replete with euphemisms and metaphors that obscured power relations or exaggerated precision---e.g., ``clean coal'', ``precision bombing'', ``scarcity'', ``survival of the fittest''. In each case, evocative language shapes public perception while masking underlying complexity, uncertainty, or harm. What distinguishes AI is therefore not a difference in kind, but a difference in scale and speed. AI discourse unfolds under conditions of capitalist acceleration, compressed publication cycles, and industry-dominated authorship networks \citep{Abdalla-Abdalla-2021}. Neologisms emerge, stabilise, and institutionalise at a pace rarely seen in other scientific domains. Once entrenched, they become difficult to dislodge, even when their conceptual shortcomings are widely recognised.

    Taken together, these dynamics explain why glosslighting is not an aberration but a stable feature of contemporary AI discourse. Anthropomorphic language fuels hype; hype attracts funding, attention, and institutional legitimacy; and institutional incentives, in turn, reward the continued use of that language. The result is a self-reinforcing cycle in which ambiguity is not merely tolerated but structurally advantageous. However, as AI systems become more deeply embedded in social, economic, and political infrastructures, the stakes of this cycle increase. What once functioned as a loose metaphor now shapes expectations about responsibility, capability, and risk. Understanding glosslighting, therefore, requires situating linguistic practices within the broader political economy of AI.

    In light of these dynamics, it is worth noting that {\it foreseeability} is a key feature of our description of glosslighting. In this case, ``foreseeable'' need not imply intention to manipulate; instead, it involves a choice to introduce additional meaning (hence, generating polysemy) to otherwise common terms instead of, e.g., coining a new term that lacks such connotation. For example, ``hallucination'' (from the Latin {\it alucinari}, meaning ``to wander in the mind'') may vary slightly in clinical and colloquial usage, but both aim to describe a family of related phenomena---roughly, the appearance of perception in the absence of stimulus. Hence, when ``hallucination'' is re-coined by computer scientists to describe a type of model output, this involves a {\it decision} to use an entrenched, colloquial term in a narrow, technical sense, regardless of whether that decision was malicious.%
        \footnote{Popular narratives attribute the introduction of the term ``hallucination'' in the AI lexicon to a 2015 blog post written by a founding member of OpenAI \citep{Karpathy-2015}. However, the term was multiply introduced to computer science much earlier, and it was used in distinct ways depending upon the sub-field in question. In text processing, as early as $1982$, ``hallucinations'' were described as ``complete mis-analyses'' created without ``any indication'' that they are inaccurate \citep{Tait-1982}. This is similar to how ``hallucination'' is used in describing LLM-outputs today, but instead of seeing these outputs as exceptional or fixable, they were seen as limiting the robustness of the system that produced them---to the extent that ``It would probably be better if the system produced no analysis at all'' \citep{Tait-1982}. In contrast, ``hallucinations'' in image-processing were described in $1985$ and $1999$ as a positive feature of the algorithm in question---in the former case \citep{Mjolsness-1985}, the ``hallucination'' was an intermediary output whereas in the latter it was the final output \citep{Baker-Kanade-2000}. Unlike today's usage, the term ``hallucination'' was co-opted in image-processing specifically to connote literally false, but sometimes useful outputs. See further discussion in \citet{Pearson-2024, Maleki-et-al-2024, Emsley-2023}.}
    As such, those who introduce glosslighting terms are still responsible for these terms being misleading because the consequences of these linguistic choices are obvious, predictable, and repeatedly realised. When glosslighting terms are deployed in contexts where non-expert, anthropomorphic interpretation is not just possible but virtually {\it guaranteed}, continued reliance on such language cannot plausibly be treated as innocent, accidental, or neutral. In this case, even if a researcher or institution claims that they do not {\it intend} audiences to attribute, e.g., literal agency, understanding, or intelligence to AI systems, the widespread and well-documented tendency to do exactly that makes such interpretations near-inevitable. Persisting in the use of misleading terminology under these conditions amounts to a form of wilful disregard. Hence, researchers, collectively, should be held responsible for these foreseen (even if unintended) consequences of their actions when those consequences are sufficiently serious---and also avoidable \citep{sep-double-effect}. The inclusion of scare quotes or disavowing footnotes does not discharge this responsibility, in the same way that the inclusion of an ethics section or high-level disclaimer about the normative implications of one's work (now mandatory at many leading conferences) serves only to signal responsibility while leaving the substantive drivers of harm untouched.

    Seen in this light, glosslighting has consequences that extend well beyond mere misinformation. Anthropomorphic framings inherent to glosslighting terms do not simply exaggerate machine capabilities; instead, they reconfigure normative benchmarks for intelligence, competence, and worth. In doing so, they risk rehabilitating and justifying hierarchies that have historically been used for exclusion, discrimination, and violence.%
        \footnote{For example, \citet{Bender-Hanna-2025} suggest that ``when we imbue these systems with fictitious consciousness, we are implicitly devaluing what it means to be human, and endorsing a much longer line of thinking about the nature of intelligence based on eugenics and race science'' (22--23).} 
    Recent work in AI underscores the re-emergence of hereditarian or eugenics-adjacent claims in computational contexts \citep{Guenzel-et-al-2025}. This further illustrates how technical framing can lend scientific legitimacy to ideas with long and well-documented histories of harm,%
        \footnote{See discussion and further examples in \citet{andrews2024reanimation, Stark-Hutson-2022, ONeil-2016}. See also \citet{Gebru-Torres-2024}.} %
    even when authors disavow those histories explicitly.

    These dynamics are especially consequential when glosslighting intersects with automation in high-stakes institutional contexts, where linguistic framing shapes how automated systems are trusted, legitimised, and acted upon. In such settings, automation is frequently presented as a neutral or corrective intervention---an appeal that is reinforced by anthropomorphic and technical-sounding language---even when it is layered over top of deeply inequitable social arrangements. The function of automation, in many of these cases, it not to support human decision-making, but to provide a false sheen of objectivity over a brutally discriminatory system \citep{Bender-Hanna-2025}. Glosslighting terms reinforce this appearance of neutrality by entrenching a veneer of technical legitimacy or objectivity, encouraging deference to algorithmic outputs while obscuring the human decisions, values, and power relations embedded within them. For example, \citet{Miceli-et-al-2025} highlight how ``bias'' (a glosslighting term) treats discrimination as a technical glitch, obscuring deeper societal issues like historical inequity, exploitative data labour, and whose world-views dominate. They suggest that a power-aware lens reveals why certain definitions and descriptions become normal and others are excluded. Our point is that glosslighting terms exacerbate the problem by making systems appear more capable, adaptive, or intelligent that they are, while crowding out the slower, more rigorous processes required to assess whether a given technology is appropriate for a particular context.

    Glosslighting, therefore, functions not merely as a linguistic phenomenon but as a {\it mechanism} that enables and exacerbates institutional harm. By amplifying confidence while diffusing accountability, it accelerates development without commensurate evaluations, legitimises harmful applications under the guise of innovation, and narrows the space of meaningful critique. The ethical issue, then, is not simply that language can sometimes mislead; it can do so in ways that are systematically advantageous to powerful actors while externalising risk onto already-vulnerable populations.%
        \footnote{See, for example, \citet{Green-2021, Falbo-LaCroix-2022, Helm-Gerlek-2025, Inie-et-al-2026, abercrombie2023, cheng-etal-2025-dehumanizing}.}
    Recognising these dynamics underscores why a critique of glosslighting cannot simply consist of calls for better communication alone. It requires a deeper reckoning with how language, incentives, and historical legacies interact in shaping what kinds of AI systems are built, deployed, and trusted, and at whose expense.

\section{Conclusion}
    \label{sec:Conclusion}

    The analysis in this paper has turned on a central distinction between {\it innocent} and {\it strategic} polysemy. While ordinary ambiguity is an ineliminable and often productive feature of natural language, strategic polysemy exploits that flexibility. Although strategic polysemy can be normatively neutral, in contemporary AI discourse a type of strategic polysemy proliferates, which we call {\it glosslighting}. Glosslighting terms, we have demonstrated, invite anthropomorphic or value-laden interpretations while preserving plausible deniability through narrower technical descriptions. Hence, this is a systematic linguistic practice that trades on familiar concepts to generate confidence, excitement, hype, and legitimacy, without incurring corresponding epistemic or normative commitments.

    We have argued that glosslighting is not merely a communicative flaw, nor is it incidental. Instead, it is a constitutive driver of the modern AI hype cycle (and it has existed, in some form or another, since the start of AI research). That said, anthropomorphic terminology does not simply accompany exaggerated expectations; it actively produces them by shaping how systems are imagined, evaluated, and governed. In doing so, glosslighting distorts epistemic norms: it obscures the probabilistic, brittle, and tool-like character of AI systems; diffuses responsibility when harms occur; and licenses speculative narratives—both utopian and catastrophic---that crowd out careful assessment. These distortions are reinforced by institutional incentives across research, industry, media, and investment, making ambiguity not an accidental by-product but a structurally advantageous feature of the AI ecosystem.

    Hence, it is worth highlighting that the more things change, the more they stay the same. In the 1970s Drew McDermott criticised the use of ``wishful mnemonics'', like {\sc goal} and {\sc understand}, to refer to programs and data structures as theoretically question-begging. This convention, \citet{McDermott-1976} suggested, ``may mislead a lot of people, most prominently [the programmer], and enrage a lot of others'' (4). Instead, he suggests, the programmer should use literal, non-anthropomorphic, or even meaningless strings of symbols, like {\sc G0034}, to name the parts of a programme and then ``see if he can {\it convince} himself or anyone else that [it] implements some part of understanding'' \citep[p. 4]{McDermott-1976}. Sometimes, anthropomorphism in the context of AI is question-begging; other times it is simply fiction \citep{Proudfoot-2011}. McDermott sought to eliminate the use of overly optimistic or wishful mnemonics, stating, ``If we are to retain any credibility, this should stop'' \citep{McDermott-1976}. He advised, as we have a half-century later, that the disciplined AI researcher should instead adopt ``colourless'' or ``sanitized'' language---terms that carry a modest, purely technical meaning.

    The normative takeaway is simultaneously straightforward and demanding. Linguistic conventions in AI are an ethical and political practice that shapes knowledge, trust, and power. Responsible scientific communication requires more than disclaimers or scare quotes---it demands the resistance of using strategically ambiguous terminology when its foreseeable uptake will mislead non-expert audiences. Linguistic transparency, therefore, is not solely a matter of pedantry but a precondition for epistemic justice in the age of AI. Without it, those who control the language of AI also control how its risks, benefits, and failures are understood, and by whom.

\begin{acks}
    The authors would like to thank each other, for their hard work and dedication.
\end{acks}

\section*{Generative AI Usage Statement}

The author(s) did not use generative AI in the writing of this paper.

\bibliographystyle{ACM-Reference-Format}
\bibliography{Biblio}

\appendix

\section{Translation Manual: Communicating AI Without Glosslighting}
    \label{sec:Appendix}

    Researchers increasingly find themselves communicating about AI systems not only to other specialists but also to journalists, policymakers, civil society organisations, and the broader public. In these contexts, the terminology used within technical communities can easily be interpreted through everyday meanings that differ substantially from their operational definitions. As discussed in the main text, such gaps between intuitive and technical meanings create fertile conditions for what we call {\it glosslighting}: the use of technically redefined or polysemous terms that evoke familiar interpretations while retaining the ability to retreat to narrower technical descriptions.

    This appendix provides a practical resource for researchers seeking to reduce such misunderstandings when engaging with media and policy audiences. The goal is not to prescribe rigid terminology, but to highlight how commonly used expressions in AI discourse may invite anthropomorphic or inflated interpretations, and to offer clearer alternatives that describe the underlying computational processes more directly.
    
    In this respect, the table below extends prior efforts to de-anthropomorphise AI discourse within the research community.%
        \footnote{For example, \citet{Rehak-2025} critically examines the narratives through which AI capabilities are framed in public discourse. Similarly, \citet{Inie-et-al-2026} recommend moving from anthropomorphic ``wishful mnemonics'' toward terminology that more accurately reflects the mechanisms of machine learning systems, based on the principal of ``functionality first''. Our contribution here is to operationalise these insights as a translation aid that researchers can use when preparing press briefings, policy testimony, public reports, or interdisciplinary collaborations.}
    Table~\ref{tab:Translation} should therefore be understood as a communication aid rather than a prescriptive glossary. In many technical contexts, established shorthand terms may be unavoidable or useful. However, when communicating outside specialist audiences, substituting or clarifying such terms can help reduce the risk that linguistic shortcuts inadvertently reinforce hype, obscure limitations, or create misleading impressions about the nature of AI systems. By making the computational mechanisms more explicit, researchers can contribute to clearer public understanding and more grounded policy discussions about AI capabilities and risks.

\begin{longtable}{p{0.3\columnwidth}p{0.65\columnwidth}}
\caption{}
\label{tab:Translation}\\
\toprule
{\bf Hype-Laden Term} & {\bf Literal Description} \\  
\midrule
    Affective AI & Sentiment-conditioned response templates \\
\midrule
    Agent & Model wrapped in a control loop with memory, tools, and a task-selection policy \\
\midrule
    Alignment & How closely the model's output distribution matches developer or user preferences, often operationalised through dataset curation, loss functions, reinforcement learning signals, or safety constraints. Could also be more simply described as a software bug or design defect. \\
\midrule
    Attention & Weighted similarity operations \\
\midrule
    Chain of Thought & Intermediate natural-language token sequences generated to improve next-token prediction \\
\midrule
    Emergence / Emergent Behaviour & Nonlinear scaling phenomena \\
\midrule 
    Empathy / Empathic AI & See {\it affective AI} \\ 
\midrule
    Friendliness & Alignment constraints or guardrails\\
\midrule
    Hallucination & Output divergence from training distribution \\ 
\midrule
    Honesty & Reliability under specific benchmarks \\
\midrule
    Inference & The process of executing a trained model on an input---i.e., carrying out the mathematical operations defined by its architecture and parameters---to produce an output. Executing a learned function once. \\
\midrule
    Intelligence & A contested and loosely defined concept; in machine learning contexts it typically refers to performance on specific benchmarks or tasks rather than general cognitive ability. \\
\midrule
    Introspection & When a model is prompted to output statements about its own internal processes or states. It does not access such states; it simply generates outputs that resemble descriptions of internal behaviour, based on patterns it learned during training. \\
\midrule 
    Knowledge & Statistical association or memorised text \\
\midrule
    Learning & Parameter updates or inference-time adaptation\\
\midrule
    Moral reasoning & pattern-matching ethical training examples \\
\midrule
    Personality & Fine-tuned output style or roleplay \\ 
\midrule
    Reasoning & Multi-step pattern completion over learned representations; heuristic search through statistically associated transformations \\
\midrule
    Scheming & Goal-conditioned text produced under certain prompts or incentives \\ 
\midrule 
    Superintelligence & Speculative (non-existent) scaling trajectory \\
\midrule
    Trustworthiness & See {\it Honesty} \\
\bottomrule
\end{longtable}

\end{document}